\documentstyle[11pt,epsfig]{article}

\def\laq{\raise 0.4ex\hbox{$<$}\kern -0.8em\lower 0.62
ex\hbox{$\sim$}}
\def\gaq{\raise 0.4ex\hbox{$>$}\kern -0.7em\lower 0.62
ex\hbox{$\sim$}}

\def\vp{\varphi}
\def \vpb {{\overline {\vp}}}
\begin{document}

\begin{titlepage}
\begin{flushright}
CERN-PH-TH/2005-013
\end{flushright}

\vspace{0.8cm}

\begin{center}

\huge{Curvature perturbations \\
from dimensional decoupling}

\vspace{0.8cm}

\large{Massimo Giovannini \footnote{e-mail address: massimo.giovannini@cern.ch}}

\normalsize

\vspace{0.3cm}
{{\sl Department of Physics, Theory Division, CERN, 1211 Geneva 23, Switzerland}}

\vspace*{1cm}

\begin{abstract}
\noindent
The scalar modes of the geometry induced by  dimensional decoupling
are investigated. In the context of the low energy string effective 
action, solutions can be found where the spatial part of the background geometry is the direct product of two maximally symmetric Euclidean manifolds whose related scale factors evolve at a dual rate so that the expanding dimensions first accelerate and then decelerate while the internal dimensions always contract. 
After introducing the perturbative treatment of the inhomogeneities,
a class of five-dimensional geometries is discussed in detail. 
Quasi-normal modes of the system are derived and the numerical solution for the evolution of the metric inhomogeneities shows that the fluctuations of the internal dimensions provide a term that can be interpreted, in analogy with the well-known four-dimensional situation, as a non-adiabatic pressure density variation. Implications of this result are discussed with particular attention to string cosmological scenarios.
\end{abstract}
\end{center}

\end{titlepage}

\newpage
\renewcommand{\theequation}{1.\arabic{equation}}
\section{Introduction and motivation(s)}
\setcounter{equation}{0}

It is not unreasonable that internal dimensions 
of Planckian size can play a r\^ole in the early 
stages of the life of the Universe \cite{appelquist}. 
The question is then how internal dimensions 
became so small in comparison with the external 
(expanding) dimensions. In the last 
twenty years various sorts of cosmological compactifications 
have been proposed \cite{shadev,abbott,kolb,astro} (see 
also \cite{appelquist} for a review).
 These mechanisms invoke, in some way 
or in another, a long phase during which the scale 
factor of the internal manifold shrinks until it stabilizes 
to the wanted Planckian value.  
There are at least three independent problems 
associated with a mechanism of cosmological conpactification:
\begin{itemize}
\item{} the nature of the sources driving the contraction 
of the internal dimensions;
\item{} the nature of the mechanism stabilizing the radius of the internal manifold;
\item{} the properties of the inhomogeneities  induced by the process of compactification.
\end{itemize}

Even higher-dimensional
Einstein equations allow for vacuum solutions of Kasner type. These 
solutions may well describe the contraction of the internal dimensions.
The caveat with such a  simple realization of cosmological compactification 
is the possible occurrence of physical singularities both in the
external and internal manifold, as repeatedly noticed 
in connection with string cosmological models \cite{astro}.  

The properties of metric perturbations induced by a phase of compactification,
have been  partially discussed by various authors. In particular, the Bardeen 
formalism \cite{bardeen} has been generalized to the case of anisotropic 
manifolds  \cite{tomita,ST}(see also \cite{DEM}). The evolution equations for the Bardeen potentials 
have been solved in a number of explicit examples \cite{den,abe}.  Canonical normal 
modes of the fluctuations have been obtained in specific classes of anisotropic manifolds \cite{ST,normal}.  

The aim of the present  paper is to argue that the contraction of the 
internal manifold is not a sufficient 
condition for the  perturbative treatment of the curvature 
inhomogeneities induced by the process of compactification. 
In the purely four-dimensional case, curvature 
fluctuations are conserved over wavelengths larger than the Hubble radius
in a number of relevant cases like, for instance, the one where 
a single scalar field sources the dynamics of the background geometry.
If internal dimensions are present, their fluctuations 
will necessarily  induce a term analog to the non-adiabatic pressure 
density variation. Therefore, curvature fluctuations will no longer be conserved. This occurrence would not be, by itself, problematic. However, a problem may arise 
if the growth of curvature fluctuations is too sharp when 
the relevant wavelengths are larger than the Hubble radius.

To discuss this problem we will use, as a toy model, the low-energy 
string effective action typical of pre-big bang models supplemented 
by a (non-local) dilaton potential. Within this set-up is possible 
to obtain rather reasonable solutions describing the compactification
of the internal dimensions and the simultaneous expansion 
of the external dimensions \cite{MV1,MV2,GMV,heat}. 
An interesting feature of these 
solutions is that the external dimensions expands 
in an accelerated fashion and then decelerate (but always expand). 
The transition from acceleration to deceleration is regular 
in a technical sense. Finally, these solutions can be usefully 
expressed in analytical terms and this aspect helps 
an unambiguous analytical treatment of the evolution of the fluctuations of the geometry.

The present paper is then organized as follows.
In Section 2 the background evolution will be 
specifically described in the case of multidimensional 
cosmological models characterized by $d$ expanding 
and $n$ contracting (internal) dimensions. In Section 3, the salient features 
of the evolution of the fluctuations of the geometry will be 
derived. Owing to the large number of scalar degrees of freedom 
present in higher-dimensional (anisotropic) manifolds, the analysis 
will be then preformed in a five-dimensional geometry. In Section 4 the 
evolution equation of the generalized curvature fluctuations 
will be presented. Moreover,  the evolution equations 
of the system will be recast in a useful first-order form.
In Section 5 the issue of quantum-mechanical normalization 
of the fluctuations will be discussed and quasi-normal modes of the system 
introduced. Section 6 collects some numerical discussion while the concluding 
remarks will be presented in Section 7. The Appendix 
contains a series of useful technical details on the 
derivation of the perturbed five-dimensional equations.

\renewcommand{\theequation}{2.\arabic{equation}}
\section{Dimensional decoupling}
\setcounter{equation}{0}

Multidimensional cosmological models 
whose line element can be expressed as
\begin{equation}
d s^2 = G_{A B} d x^{A} d x^{B} = dt^2 - a^2(t)\delta_{ij} dx^i d x^{j} - 
b^2(t) \delta_{ab} dy^{a} dy^{b},
\label{geometry}
\end{equation}
will be considered in the present Section. In Eq. (\ref{geometry})
 $G_{AB}$ is the full metric; the indices $A,B$ run over the full $D = d + n + 1$-dimensional
space-time; the $d$-expanding dimensions are typically three 
and the $n$-contracting dimensions are generic but the case $n=1$ will be of particular interest; for notational 
convenience, in explicit formulae, the indices $i,j,k,l...$ 
run over the $d$ expanding dimensions and the 
indices $a,b,c,d...$ run over the $n$ internal dimensions.  

The background equations stemming from the low-energy string effective 
action can then be written, in natural string units, as  \cite{astro}
\begin{eqnarray}
&& \dot{\vpb}^2  - d H^2 - n F^2 - V= 0,
\label{b1}\\
&& \dot{H} = H \dot{\vpb},
\label{b2}\\
&&  \dot{F} = F \dot{\vpb},
\label{b3}\\
&& 2 \ddot{\vpb} - \dot{\vpb}^2 - d H^2 - nF^2 + V - \frac{\partial V}{\partial\vpb}=0,
\label{b4}
\end{eqnarray}
where $H = \dot{a}/a$ and $F= \dot{b}/b$; $\overline{\varphi} = \varphi - \log{\sqrt{-G}}$
is the so-called shifted dilaton; the overdot denotes a 
derivation with respect to the cosmic time coordinate $t$. 
The potential $V$ is only function of $\overline{\varphi}$, i.e. 
$ V= V(\overline{\varphi})$. Eqs. (\ref{b1})--(\ref{b4}) can be derived 
from the following (generally covariant) action 
\footnote{The choice of natural string units implies that $ 2 \lambda_{\rm s}^{D -2} =1$.}
\begin{equation}
S= - \frac{1}{2\lambda_{s}^{D-2} }\int d^{D-1} x \sqrt{|G|} e^{- \varphi} \biggl[ R + 
G^{\alpha\beta} \nabla_{\alpha} \varphi \nabla_{\beta} \varphi + V(\overline{ \varphi})\biggr],
\label{action1}
\end{equation}
where 
\begin{eqnarray}
&& V= V(e^{-\overline{\varphi}}),
\nonumber\\
&& e^{-\overline{\varphi}(x)} = \frac{1}{\lambda_{s}^{D-1}} \int d^{D} w \sqrt{|G(w)|} e^{- \varphi(w)} 
\sqrt{G^{\alpha\beta} \partial_{\alpha} \varphi (w)\partial_{\beta} \varphi(w)} \delta\biggl( \varphi(x) - \varphi(w)\biggr).
\end{eqnarray}

The variation of the action (\ref{action1}) with respect to $G_{\mu\nu}$ and $ \varphi$ leads, after linear combination,  to Eqs. (\ref{b1})--(\ref{b4}) 
valid in the case of homogeneous dilaton field. The general ($D$-dimensional)  derivation of Eqs. (\ref{b1})--(\ref{b4}) from Eq. (\ref{action1}) has 
been performed in Ref. \cite{max2} (see also \cite{max1}) and will not be repeated here.
The combination of Eqs. (\ref{b1}) and (\ref{b4}) and of 
Eqs. (\ref{b2})--(\ref{b3}) leads, respectively, to the following pair of 
equations
\begin{eqnarray}
&& \dot{\overline{\varphi}}^2 - \ddot{\overline{\varphi}} 
+ \frac{1}{2} \frac{\partial V}{\partial \overline{\varphi}} - V =0,
\label{B1}\\
&& (\dot{H} - \dot{F}) = (H - F) \dot{\overline{\varphi}}.
\label{B2}
\end{eqnarray}

Using Eqs. (\ref{B1}) and (\ref{B2}) it is rather easy to find solutions of the 
whole system for various sets of potentials. Consider, in particular, the case 
of exponential potential $ V(\overline{\varphi} ) = - V_{0}e^{4\overline{\varphi}}$.
In this case the solution of the system can be expressed as 
\begin{eqnarray}
&& a(t) = (t + \sqrt{t^2 + t_{0}^2})^{\frac{1}{\sqrt{d + n}}},
\label{as1}\\
&& b(t) =  (t + \sqrt{t^2 + t_{0}^2})^{-\frac{1}{\sqrt{d + n}}},
\label{bs1}\\
&& \overline{\varphi} = \varphi_{0} - \frac{1}{2} \ln{(t^2 + t_{0}^2)}.
\label{phs1}
\end{eqnarray}
To satisfy the system (\ref{b1})--(\ref{b4}), the free parameters 
of the solution are subjected to the consistency relation 
$e^{4 \varphi_{0}}\,\,\, t_{0}^2\,\,\, V_{0}\,\,\, =1$.
Concerning the above solution few comments are in order:
\begin{itemize}
\item{} the solutions are defined for $ t$ between $ -\infty$ and $+\infty$;
\item{} the solutions are regular (in a technical sense) since 
all the curvature invariants are regular for every $t$;
\item{}the class of solutions expressed by Eqs. (\ref{as1})--(\ref{phs1}) 
describe a situation of dimensional decoupling. In fact while  
$a(t)$ is always expanding (i. e. $\dot{a} >0$), $b(t)$ is always 
contracting (i.e. $\dot{b} <0$) at a dual rate (i.e. $b(t) \sim 1/a(t)$).
\end{itemize}

In connection with the last point, it can be easily shown that $a(t)$
expands in an accelerated 
way for $t <0$ (i.e. $\ddot{a} >0$), while the expansion becomes decelerated 
for $t> 0$ (i.e. $\ddot{a}<0$). In fact, taking the limit of Eqs. (\ref{as1}) and (\ref{bs1}) 
for $t \to \pm \infty$, the leading behaviour of the scale 
factors will be 
\begin{eqnarray}
&& a(t) \sim (-t)^{- \frac{1}{\sqrt{d + n}}}\,\,\,\,\,\,\,\,b(t) \sim (-t)^{\frac{1}{\sqrt{d + n}}}, \,\,\,\,\,\,\,\,\, t<0;
\nonumber\\
&& a(t) \sim t^{ \frac{1}{\sqrt{d + n}}}\,\,\,\,\,\,\,\,b(t) \sim t^{-\frac{1}{\sqrt{d + n}}}, \,\,\,\,\,\,\,\,\, t>0.
\label{limits}
\end{eqnarray}

On the basis of Eqs. (\ref{limits}), the case of a five-dimensional geometry is particularly interesting 
since, for $t \to +\infty$ the scale factor behaves as $ a(t) \sim \sqrt{t}$, i.e. as if the Universe would 
be dominated by radiation. 

\renewcommand{\theequation}{3.\arabic{equation}}
\section{Evolution equations of the fluctuations and gauge choices}
\setcounter{equation}{0}
In higher-dimensional manifolds characterized 
by $D$ space-time dimensions, the 
perturbed metric $\delta G_{A B}$ has, overall, 
 $D( D+ 1)/2$ degrees of freedom.  Since 
 $D$ gauge conditions can be imposed 
 to fix the origin of the coordinate system, the 
 total number of degrees of freedom becomes, after
 gauge-fixing, $D(D-1)/2$.  If $D=4$, it is rather 
 obvious to classify fluctuations depending on their 
 transformation properties under three-dimensional 
 rotations. This is the choice usually employed in the 
 context of the Bardeen formalism. If $ D > 4$ one 
 can, in principle, classify the metric fluctuations 
 according to their transformation properties 
 under some larger group of symmetries. If 
 the spatial part of the background geometry is decomposed as in Eq. (\ref{geometry}), it is practical 
 to classify fluctuations with respect 
 to the rotations in the expanding dimensions. This 
 was at least a possibility studied in \cite{ST,normal}
 (see also \cite{tomita}). In this approach, 
as soon as the number of internal dimensions increases 
the number of scalar modes also increases.
For practical reasons, the analysis will then 
 be conducted in the case of a five-dimensional 
geometry. Extensions of the formalism to $D>5$ 
have been also proposed \cite{ST,normal}.

To derive  the evolution of the fluctuations 
of the geometry it is 
practical to work directly in the conformal time parametrization, i.e. 
$a(\eta) d \eta = dt$. Within this parametrization the fluctuations 
of a five-dimensional geometry can be written as 
\begin{eqnarray}
&&\delta G_{00} = 2 a^2 \phi,
\nonumber\\
&& \delta G_{0 i} = - a^2 \partial_{i} P - a^2 Q_{i},
\nonumber\\
&& \delta G_{0 y} = - a b C,
\nonumber\\
&&\delta G_{i j} = 2 a^2 \psi \delta_{i j} - 2 a^2 \partial_{i} \partial_{j} E + a^2 h_{i j}+ a^2 (\partial_{i} W_{j} + \partial_{j} W_{i}), 
\nonumber\\
&& \delta G_{i y} = - a b \partial_{i} D - a b H_{i},
\nonumber\\
&& \delta G_{yy} = 2 b^2 \xi,
\label{param}
\end{eqnarray}
where the tensor (i.e. $h_{ij}$) and vector (i.e. $ W_{i}$, $Q_{i}$ and $H_{i}$)
modes satisfy:
\begin{eqnarray}
&& \partial_{j} h_{i}^{j}=0,~~~~h_{i}^{i} =0,
\nonumber\\
&& \partial_{i} W^{i}=0,~~~~~~\partial_{i} Q^{i}=0,~~~~\partial_{i}H^{i} =0. 
\end{eqnarray}
So, $h_{ij}$ carries 2 degrees of freedom; $W_{i}$, $Q_{i}$ and $H_{i}$ carry overall 6 degrees of freedom; and we also have seven scalars. This implies, as expected, that the parametrization (\ref{param}) 
respects the total number of degrees of freedom of the perturbed five-dimensional metric. For infinitesimal coordinate transformations
\begin{equation}
x^{A}\to \tilde{x}^{A} = x^{A} + \epsilon^{A},
\end{equation}
with gauge functions
\begin{equation}
\epsilon^{A} =( \epsilon^{0}, \epsilon^{i}, \epsilon^{y}),~~~~~~~
\epsilon_{A} =(a^2 \epsilon_{0},-a^2 \epsilon_{i}, -b^2\epsilon_{y}),
\end{equation}
the tensor modes of the 
geometry are invariant while the vectors and scalars transform, 
respectively, as \footnote{The prime denotes a derivation with 
respect to the conformal time coordinate while the dot denotes, as 
previously reminded, the derivation with respect to the 
cosmic time coordinate. The relation between the two derivatives 
is simply $\partial/\partial\eta = a \partial/\partial t$.}
\begin{eqnarray}
 && \tilde{Q}_{i} = Q_{i} - \zeta_{i}',
 \nonumber\\
&& \tilde{W}_{i}= W_{i} + \zeta_{i},
\nonumber\\
&&\tilde{H}_{i} = H_{i} - \frac{a}{b} \partial_{y} \zeta_{i}, 
\end{eqnarray}
and 
\begin{eqnarray}
&& \tilde{\phi} = \phi - \epsilon_0' - {\cal H} \epsilon_0,
\nonumber\\
&& \tilde{P} = P - \epsilon' + \epsilon_0,
\nonumber\\
&&\tilde{C} = C - \frac{b}{a} \epsilon_{y}' + \frac{a}{b} \partial_{y} \epsilon_{0},
\nonumber\\
&& \tilde{\psi} = \psi + {\cal H}\epsilon_0,
\nonumber\\
&& \tilde{E} = E - \epsilon,
\nonumber\\
&& \tilde{D} = D - \frac{b}{a} \epsilon_{y} - \frac{a}{b} \partial_{y} \epsilon,
\nonumber\\
&& \tilde{\xi} = \xi + \partial_{y} \epsilon_{y} + {\cal F} \epsilon_0,
\end{eqnarray}
where the following notation has been used 
\begin{equation}
\epsilon_{i} = \partial_{i} \epsilon + \zeta_{i}
\end{equation}
such that $\partial_{i} \zeta^{i} =0$.
On top of the fluctuations of the geometry, the 
fluctuations of the sources should also be taken into account:
the gauge transformation for the dilaton fluctuation is then 
\begin{equation}
\tilde{\chi} = \chi - \varphi' \epsilon_{0} 
\end{equation}
where $\delta \varphi = \chi$. 
For instance, the longitudinal gauge \cite{bardeen} 
can be generalized to the higher-dimensional case. For the applications 
under study it is practical to a adopt the generalized 
uniform dilaton gauge stipulating that
\begin{equation}
\tilde{W}_{i} =0,~~~~~\tilde{D}=0, ~~~~~\tilde{P}=0, ~~~~~\tilde{\chi}=0.
\label{gaugecond}
\end{equation}
that has been already exploited in various four-dimensional calculations \cite{max2,max1}.  Notice 
that Eq. (\ref{gaugecond}) corresponds to five indipendent conditions 
since $\tilde{W}_{i}$ is divergenceless. Different gauge 
choices can be adopted (see, for instance, \cite{ST}).
However, it has been previously shown \cite{max1} that these
gauge choices are less suitable, already in four dimensions, since they lead  to divergences in the 
evolution equations of the fluctuations. The expectation 
that the (generalized) uniform dilaton gauge 
leads to non-singular evolution equations will be confirmed 
by the present analysis.

In the uniform dilaton gauge the perturbed equations can be written in explicit 
terms (the details of the derivation are reported in the Appendix).
The perturbed form of the generally covariant equations 
can be written as \footnote{Following common practice,
$\overline{G}_{M N}$ and $\overline{\Gamma}_{A B}^{C}$
denote the background values of the metric and of the 
Christoffel connections, while $\delta G_{M N}$ and 
$\delta \Gamma_{A B}^{C}$ are the first-order 
fluctuations of the corresponding quantities.}
\begin{eqnarray}
&& \delta {\cal G}_{A}^{B} + \delta G^{B M} \biggl[ \partial_{M}\partial_{A} \vp - \overline{\Gamma}_{M A}^{N} \partial_{N} \vp\biggr]
- \overline{G}^{B M} \delta \Gamma_{M A}^{N} \partial_{N} \vp 
\nonumber\\
&& +\frac{1}{2} \delta_{A}^{B} \biggl[ \delta G^{M N} \partial_{M}\vp \partial_{N} \vp -  2 \delta G^{M N} (  \partial_{M} \partial_{N}\vp - 
\overline{\Gamma}_{M N}^{C} \partial_{C}\vp)
+ 2 \overline{G}^{M N} \delta \Gamma_{M N}^{C} \partial_{C} \vp\biggl]
\nonumber\\
&& - \frac{1}{2} \frac{\partial V}{\partial \overline{\vp}} \phi \overline{\gamma}_{A}^{B}  =0,
\label{PERT2}
\end{eqnarray}
where $\delta {\cal G}_{A}^{B}$ denotes the fluctuation of the Einstein 
tensor and where
\begin{equation}
\overline{\gamma}_{A}^{B} 
= \delta_{A}^{B} - \frac{\partial^{B} \varphi \partial_{A} \varphi}{\overline{G}^{M N}\partial_{M}\varphi \partial_{N} \varphi}.
\end{equation}
Equation (\ref{PERT2}) holds in the case of the scalar modes of the geometry
since terms like $\delta \gamma^{i}_{0} \propto Q^{i}$, relevant only 
in the analysis of the vector modes \cite{max3}, have been omitted.  Further details on the derivation of the 
evolution equations of the first-order fluctuations
are reported in the Appendix (see, in particular, 
Eqs. (\ref{EINT}) and (\ref{EOM})).

The components of Eq. (\ref{PERT2}), computed 
explicitly in Eqs. 
(\ref{EOM}), 
 give the  set of 
perturbed equations for the evolution of the fluctuations.
More specifically, from the perturbed $(00)$ 
and $(0i)$  components of Eq. (\ref{PERT2}), the Hamiltonian 
and momentum  constraints are, respectively,
\begin{eqnarray}
&& \phi[ {\cal F}^2 - {{\vpb}'}^2 + 3 {\cal H}^2] + ({\vpb}' + {\cal F}) \xi' 
+  3 ({\vpb}' + {\cal H}) \psi'  
\nonumber\\
&&+ \frac{a}{b}({\vpb}' + {\cal F})\partial_{y}C + 3 \frac{a^2}{b^2} \nabla^2_{y} \psi + 
\nabla^2[\xi + 2 \psi - ({\vpb}' + {\cal H}) E' - \frac{a^2}{b^2} \nabla^2_{y} E]=0,
\label{00eq}
\end{eqnarray}
and 
\begin{equation}
-\partial_{i}[ 2\psi' + \xi' + ({\cal F} - {\cal H}) \xi - ({\vpb}' + {\cal H})\phi + 
\frac{a}{2b}\partial_{y} C ]=0.
\label{0ieq}
\end{equation}
In Eqs. (\ref{00eq})--(\ref{0ieq}) and in the following 
equations $\nabla^2_{y}$ denote the Laplacian with respect 
to the internal coordinate $y$.

From the  $(i=j)$ component of Eq. (\ref{PERT2}) we have 
\begin{eqnarray}
&& \phi[ 2 {\vpb}'' - {{\vpb}'}^2 + 2  {\cal H}' - 5 {\cal H}^2 
- {\cal F}^2 - 4 {\cal H} {\vpb}' 
- \frac{a^2}{2} \frac{\partial V}{\partial \vpb}]
\nonumber\\
&& - 2 \psi'' - \xi'' + ({\vpb}' + {\cal H})\phi' + 2 ({\vpb}' +  {\cal H})\psi ' + 
({\vpb}' + 2 {\cal H} - {\cal F}) \xi' 
\nonumber\\
&& + \nabla^2[ E'' - ({\vpb}' + {\cal H}) E' + \psi + \xi - \phi 
- \frac{a^2}{b^2} \nabla^2_{y} E] 
\nonumber\\
&&+ \frac{a^2}{b^2}\nabla^2_{y}( 2 \psi - \phi) - \frac{a}{b} \partial_{y}[ C' - 
({\vpb}' + {\cal H}) C]=0.
\label{ijeq}
\end{eqnarray}
Finally the  $(i \neq j)$, $(yy)$, $(0y)$ and $(iy)$ 
components of Eq. (\ref{PERT2}) are, respectively,  
\begin{eqnarray}
&&\partial_{i}\partial^{j}[  E'' - ({\vpb}' + {\cal H}) E' + \psi + \xi - \phi - \frac{a^2}{b^2} \nabla^2_{y} E] =0,
\label{ineqj}\\
&&- 3 \psi'' + ({\vpb}' + {\cal F})(\phi' + 3 \psi') 
\nonumber\\
&&+
\phi[ 2 {\vpb}'' - {{\vpb}'}^2 + 2 {\cal F}' - 2 ({\cal F} + {\cal H}) {\vpb}' - 
2 {\cal H} {\cal F} - 3 {\cal H}^2 - {\cal F}^2 
- \frac{a^2}{2} \frac{\partial V}{\partial\vpb}]
\nonumber\\
&&+  \nabla^2[ 2 \psi -\phi + E'' - ({\vpb}' + {\cal F})E'] =0,
\label{yyeq}\\
&&\frac{a}{2b} \nabla^2 C - \frac{a^2}{b^2} \partial^{y}\biggl\{ 3 \psi' -({\vpb}' + {\cal F})\phi +
3 ({\cal H} - {\cal F}) \psi - \nabla^2 [ E' + ({\cal H} - {\cal F})E] \biggr\}=0,
\label{0yeq}\\
&&\frac{a}{2 b} \partial_{i}[ C' - ( {\vpb}' + 2 {\cal H} - {\cal F}) C] -
\frac{a^2}{b^2} \partial_{i}\partial^{y}[ 2 \psi - \phi] =0.
\label{iyeq}
\end{eqnarray}

Concerning the above system of equations 
the following remarks  are in order:
\begin{itemize}
\item{} if the gradients with respect to $y$ are neglected, 
the evolution of $C$ decouples;
\item{} not all the equations are independent;
\item{} the equations for the zero-modes with 
respect to $y$ have a slightly more tractable form.
\end{itemize}

Let us now prove, in detail, each of the above statements.
Since Eqs. (\ref{0yeq}) and (\ref{iyeq})
imply, respectively, 
$\nabla^2 C=0$ and 
$C'= (\vpb' + 2 {\cal H} -{\cal F})C$, indeed the evolution 
of $C$ decouples.

Concerning the second remark, from
Eq. (\ref{0ieq}) the variable $\phi$ can be expressed as
\begin{equation}
\phi = \frac{2 \psi' + \xi' + ( {\cal F} - {\cal H}) \xi}{({\vpb}' + {\cal H})} + 
\frac{a}{2 b} \frac{\partial_{y}C}{({\vpb}' + {\cal H})}.
\label{int1}
\end{equation}
Furthermore, from Eq. (\ref{ineqj}), the variable $E''$ can be 
expressed as 
\begin{equation}
E'' = ( \vpb' + {\cal H}) E' + \frac{a^2}{b^2} \nabla^2_{y} E + 
\frac{2 \psi' + \xi' + ( {\cal F} - {\cal H}) \xi}{({\vpb}' + {\cal H})} + 
\frac{a}{2 b} \frac{\partial_{y}C}{({\vpb}' + {\cal H})} - \psi - \xi,
\label{int2}
\end{equation}
where the dependence upon $\phi$ has been eliminated through 
Eq. (\ref{int1}).

Equations (\ref{int1}) and (\ref{int2})  can be used to eliminate 
$\phi$ and $E''$   in   Eq. (\ref{ijeq}). The remaining dependence 
on $C$ can be simplified by means of Eq. (\ref{iyeq}). 
Then, repeated use of the  background equations imply that 
Eq. (\ref{ijeq}) is identically satisfied. 
 
It is here appropriate to recall a useful form of the 
five-dimensional
background equations in the conformal time parametrization, namely
\begin{eqnarray}
&& \overline{\varphi}'' = \overline{\varphi}' ( \overline{\varphi}' + {\cal H}) + f (V),
\label{cb1}\\
&& {\cal H}' - {\cal F}'= ({\cal H} - {\cal F}) (\overline{\varphi}' + {\cal H}),
\label{cb2}\\
&& {\overline{\varphi}'}^2 - 3 {\cal H}^2 - {\cal F}^2 = V a^2,
\label{cb4}
\end{eqnarray}
where 
\begin{equation}
f(V) = \frac{1}{2} \biggl( \frac{\partial V}{\partial \overline{\varphi}} - 2 V\biggr) a^2 
\end{equation}
depends on the specific form of the potential. 
Eqs. (\ref{cb1})--(\ref{cb4}) can be 
easily obtained from Eqs. (\ref{b1})--(\ref{b4}) and by taking a combination 
of Eqs. (\ref{b1}) and (\ref{b4}). Notice, finally, that in the case of the 
potential leading to the class of solutions described in Eqs. 
(\ref{as1})--(\ref{phs1}), $f(V) = Va^2$.

Using Eqs. (\ref{cb1})--(\ref{cb4}) and neglecting the 
gradients with respect to the 
internal coordinate, the system of the perturbed equations can be reduced 
to a more tractable form. In particular, Eqs. (\ref{00eq}) and  (\ref{0ieq})
lead, respectively, to
\begin{equation}
\nabla^2( \xi + 2 \psi) - (\overline{\varphi}' + {\cal H}) \nabla^2 E' - V a^2 \phi 
+ ( \overline{\varphi}' + {\cal F})\xi' + 3( \overline{\varphi}' + {\cal H} )\psi' =0,
\label{00I}
\end{equation}
and 
\begin{equation}
2 \psi' + \xi' + ( {\cal F} - {\cal H}) \xi - (\overline{\varphi}' + {\cal H}) \phi =0,
\label{0iI}
\end{equation}
while Eq. (\ref{ijeq}) implies
\begin{equation}
 - 2 \psi'' - \xi'' + ({\vpb}' + {\cal H})\phi' + 2 ({\vpb}' +  {\cal H})\psi ' + 
({\vpb}' + 2 {\cal H} - {\cal F}) \xi'  + f(V) \phi=0.
\label{ijI}
\end{equation}

Finally, Eqs. (\ref{ineqj}) and (\ref{yyeq}) become, respectively,
\begin{eqnarray}
&& E'' - ({\vpb}' + {\cal H}) E' + \psi + \xi - \phi =0,
\label{ineqjI}\\
&& - 3 \psi'' + ( {\vpb}' + {\cal F}) ( \phi' + 3 \psi') - f(V) \phi + \nabla^2 (\psi - \xi) 
- ({\cal F} - {\cal H}) \nabla^2 E'=0.
\label{yyI}
\end{eqnarray}

\renewcommand{\theequation}{4.\arabic{equation}}
\section{Generalized curvature fluctuations}
\setcounter{equation}{0}
For the purposes of the subsequent investigation, it is practical to define the variable 
\begin{equation}
\lambda = \psi + \frac{\xi}{2}.
\label{lambda}
\end{equation}
In terms of $\lambda$, Eq. (\ref{int1}) becomes 
\begin{equation}
\phi = \frac{2 \lambda' }{(\vpb' + {\cal H})} + 
\frac{{\cal F} - {\cal H}}{\vpb' + {\cal H}} \xi.
\label{philam}
\end{equation}
Inserting Eq. (\ref{philam}) into Eq. (\ref{00I}) and using Eqs. (\ref{cb1})--(\ref{cb4}) the generalized evolution equation for the curvature parturbations 
can be obtained:
\begin{eqnarray}
&& \lambda' = \frac{(\vpb' + 3 {\cal H} - 2 {\cal F})( \vpb' + {\cal H})}{2[( \vpb' + 3 {\cal H})^2 + 2 {\cal F}^2]} \xi' 
- \frac{ ({\cal F} - {\cal H})( 3 {\cal H}^2 + {\cal F}^2 -{\vpb'}^2)}{( \vpb' + 3 {\cal H})^2 + 2 {\cal F}^2}\xi
\nonumber\\
&& - \frac{2 ( \vpb' + {\cal H})}{( \vpb' + 3 {\cal H})^2 + 2 {\cal F}^2} \nabla^2 \lambda + 
\frac{( \vpb' + {\cal H})^2}{( \vpb' + 3 {\cal H})^2 + 2 {\cal F}^2} \nabla^2 E'.
\label{lambdaprime}
\end{eqnarray}
Clearly, in the limit ${\cal F} \to 0$ and $\xi \to 0$ (i.e. $\lambda \to \psi$), 
Eq. (\ref{lambdaprime}) reduces to 
\begin{equation}
\psi' = - \frac{2 ( \vpb' + {\cal H})}{( \vpb' + 3 {\cal H})^2} \nabla^2 \psi+ 
\frac{( \vpb' + {\cal H})^2}{( \vpb' + 3 {\cal H})^2 } \nabla^2 E',
\label{psiprime}
\end{equation}
which is the typical form for the evolution of curvature perturbations 
in the uniform dilaton gauge \cite{max1}.  Eq. (\ref{psiprime}) 
suggests that if $\varphi' = (\vpb' + 3 {\cal H}) \neq 0$ (as 
in the class of models discussed in \cite{max1}), then $\psi' \simeq 0$ 
for long wavelengths larger than the Hubble radius.
If internal dimensions are dynamical, Eq. (\ref{lambdaprime}) 
 generalizes the evolution of the curvature perturbations. 
 
Using the definition (\ref{lambda})  and 
inserting Eq. (\ref{philam}) into Eq. (\ref{ineqjI}),
 Eq. (\ref{lambdaprime})  allows to eliminate $\lambda'$,
and the resulting equation will be 
\begin{eqnarray}
&& E'' = \biggl[ ( \vpb' + {\cal H} ) E' + \frac{ 2 ( \vpb' + {\cal H})}{[ ( \vpb' + 3 {\cal H})^2 + 2 {\cal F}^2]} \nabla^2 E' \biggr] - \biggl[ \lambda + \frac{4}{[ ( \vpb' + 3 {\cal H})^2 + 2 {\cal F}^2]} \nabla^2 \lambda \biggr]
\nonumber\\
&& 
+ \frac{(\vpb' + 3 {\cal H} - 2 {\cal F})}{ ( \vpb' + 3 {\cal H})^2 + 2 {\cal F}^2} \xi' 
- \frac{[ {\vpb'}^2 + 15 {\cal H}^2 + 12 {\cal H} \vpb' + 2 
 {\cal F}^2 - 6 {\cal F} \vpb' - 6 {\cal H} {\cal F}  ]}{ 2 [ ( \vpb' + 3 {\cal H})^2 + 2 {\cal F}^2]} \xi
\label{Eprime}
 \end{eqnarray}
  Since the background evolution has a simple analytical form in $t$ 
  (but not in $\eta$), Eqs. 
 (\ref{lambdaprime}) and (\ref{Eprime}) can be written, for 
 future simplicity, in the cosmic time coordinate and in terms 
 of the auxiliary variables 
 \begin{equation}
 {\cal E} = a^2 \dot{E}, \,\,\,\,\,\,\,\,\, {\cal P} = \dot{\xi},
 \label{variablles}
 \end{equation}
 whose convenience will become clear in the context of the numerical 
 analysis. The result of this simple algebra will be, in Fourier space,
 \begin{eqnarray}
&& \dot{\lambda}_{k} =  \frac{(\dot{\vpb} + H) ( \dot{\vpb} + 3 H - 2 F)}{
2[(\dot{\vpb} + 3 H)^2 + 2 F^2]} {\cal P}_{k}
- \frac{(F- H) (3 H^2 + F^2 - \dot{\vpb}^2)}{(\dot{\vpb} + 3 H)^2 + 2 F^2}\xi_{k}
\nonumber\\
&& - \frac{  ( \dot{\vpb} + H)^2}{(\dot{\vpb} + 3 H)^2 + 2 F^2} \omega^2  {\cal E}_{k} 
+ \frac{2 (\dot{\vpb} + H)}{( \dot{\vpb} + 3 H)^2 + 2 F^2} \omega^2 \lambda_{k},
\label{cosmic1}
\end{eqnarray}
where the physical frequency of the fluctuation, i.e.  $\omega = k/a$ has 
been introduced. With similar manipulations, the evolution equation for 
${\cal E}$ will be instead:
\begin{eqnarray}
&& \dot{{\cal E}}_{k}= \biggl[ \dot{\vpb} + 2 H - 
\frac{ 2 \omega^2 ( \dot{\vpb} + H)}{(\dot{\vpb} + 3 H)^2 + 2 F^2} \biggr] 
{\cal E}_{k}+ \biggl[ \frac{\dot{\vpb} + 3 H - 2 F}{(\dot{\vpb} + 3 H)^2 + 2 F^2} \biggr] {\cal P}_{k}
\nonumber\\
&& - \frac{ [\dot{\vpb}^2 + H(15 H + 12  \dot{\vpb}) + 2 F^2 - 
6 F (\dot{\vpb} + H)]}{2[(\dot{\vpb} + 3 H)^2 + 2 F^2]} \xi_{k}
 -\biggl[ 1 - \frac{ 4 \omega^2}{(\dot{\vpb} + 3 H)^2 + 2 F^2}\biggr] \lambda_{k}.
\label{cosmic2}
\end{eqnarray}

Equations (\ref{cosmic1}) and (\ref{cosmic2}) define 
a first-order linear differential system (with time-dependent coefficients)
in the variables, ${\cal E}_{k}$, $\lambda_{k}$, ${\cal P}_{k}$ and $\xi_{k}$.
Having already specified the equation for $\dot{\xi}_{k}$, i.e. 
\begin{equation}
\dot{\xi}_{k}= {\cal P}_{k},
\label{xiP}
\end{equation}
we just need a further relation determining $\dot{{\cal P}}_{k}$. 
After some algebra (which will be swiftly mentioned in the following 
section in connection with the problem of the canonical normal modes)
the wanted evolution equation turns out to be 
\begin{eqnarray}
&& \dot{{\cal P}}_{k}= \biggl\{ \dot{\vpb} + 2 V \frac{F ( \dot{\vpb} + 3 H - 2 F)}{
(\dot{\vpb} + 3 H + F) [ (\dot{\vpb} + 3 H)^2 + 2 F^2]}\biggr\} {\cal P}_{k}
\nonumber\\
&& - 4 V \omega^2 \frac{ (\dot{\vpb} + 3 H ) ( \dot{\vpb} + 3 H - 2 F)}{( \dot{\vpb} 
+ 3 H + F)^2 [ ( \dot{\vpb} + 3 H)^2 + 2 F^2]} \lambda_{k}
\nonumber\\
&& +\biggl[ 6 V \frac{( F - H) (\dot{\vpb} + 3 H - 2 F)( 3 H^2 + F^2 + H \dot{\vpb})}
{(\dot{\vpb} + 3 H + F)^2 [(\dot{\vpb}+ 3 H)^2 + 2 F^2]}  + \omega^2 \biggr]\xi_{k} 
\nonumber\\
&& - 4 V \frac{(\dot{\vpb} + 3 H - 2 F) [ F^2 + 3 H^2 + H \dot{\vpb}]}{(\dot{\vpb} 
+ 3 H + F)^2 [ (\dot{\vpb} + 3 H)^2 + 2 F^2]} \omega^2 {\cal E}_{k},
\label{cosmic3}
\end{eqnarray}
 where the explicit form of the potential mentioned before Eq. (\ref{as1}) has 
 been used.
 
Concerning Eqs. (\ref{cosmic1}), (\ref{cosmic2}) and (\ref{cosmic3}) 
it is appropriate to recall that the time-dependent coefficients are 
always regular for any finite value of the cosmic time coordinate 
as it can be argued by noticing that, in the denominators, the only 
combinations that appear are 
\begin{equation}
(\dot{\vpb} + 3 H + F),\,\,\,\,\,\,\,\,\,\,\, (\dot{\vpb} + 3 H)^ 2+ 2 F^2.
\label{zeroes}
\end{equation}
It is clear that the second expression in Eq. (\ref{zeroes}) never goes to zero since 
it is the sum of two positive definite quantities. Moreover, 
the first quantity, corresponding to $\dot{\varphi}$ is also 
positive definite for the class of backgrounds obtained by setting
$d=3$ and $n=1$ in Eqs. (\ref{as1}) and (\ref{bs1}).

Equations (\ref{cosmic1}), (\ref{cosmic2}) and (\ref{cosmic3}) can then be 
numerically integrated once the initial conditions for the fluctuations are 
set in the limit $t \to -\infty$.  Consequently, it is  mandatory to determine 
which are the correct canonical normal modes that have to be used 
to enforce, for instance, quantum mechanical initial 
conditions for the fluctuations.

\renewcommand{\theequation}{5.\arabic{equation}}
\section{Quasi-normal modes}
\setcounter{equation}{0}

The fluctuations of the geometry have to be normalized in 
the limit $ \eta \to -\infty$ (or $ t \to - \infty$)
and, for this purpose, the canonical normal modes 
have to be determined. In the system under consideration,
when $t \to -\infty$ the potential term becomes 
negligible with respect to $ \dot{\vp}^2$, i.e. $V \ll \dot{\vp}^2$.
In this situation, two normal modes of the system can 
be identified. They will be called quasi-normal 
modes since, for $t \sim t_{0}$ (when the potential dominates)
the two modes mix non-trivially. 

As far as the {\em evolution} 
of the system is concerned, it is necessary to discuss directly 
the system derived in Eqs. (\ref{cosmic1}),(\ref{cosmic2}) 
and (\ref{cosmic3}). The reason why this choice is 
not arbitrary stems from the particular form of the 
equations characterizing the normal modes of the system.
These equations contain time-dependent coefficients 
whose denominators include powers of either $( \dot{\vpb}+ H)$ 
or $( \dot{\vpb}+ F)$. 
Now, these terms lead, generically to poles in the 
time-dependent coefficients of the system.  For instance, in the 
specific case of the class of solutions (\ref{as1})--(\ref{phs1})
the zeros of $( \dot{\vpb}+ H)$ 
and  $( \dot{\vpb}+ F)$ are, respectively, in $(d + n -1)^{1/2}$ and 
$- (d + n-1)^{-1/2}$. For the specific case of the numerical 
analysis to be presented in the following section 
this would imply poles (of various degree) for 
either $1/\sqrt{3}$ of $-1/\sqrt{3}$.

To prove the statements of the previous two paragraphs,
let us derive the evolution equations for the normal 
modes of the system first by looking at the evolution
equations and then by looking directly at the action 
perturbed to second order in the amplitude 
of the fluctuations.
Summing up Eqs. (\ref{00I}) and (\ref{ijI}) and eliminating 
$\phi$ by means of Eq. (\ref{ineqjI}) the following equation can be derived:
\begin{equation}
\nabla^2 \lambda - \lambda'' + 3 ( \vpb' + {\cal H}) \lambda' +  g( V) \phi
=-  \frac{\vpb' + {\cal H}}{2} [ E''' - (\vpb' + {\cal H}) E'' - (\vpb'' + {\cal H}') E' - 
\nabla^2 E'],
\label{lambdapp}
\end{equation}
where Eq. (\ref{lambda}) has been used and where
\begin{equation}
g(V) = \frac{a^2}{4}\biggl( \frac{\partial V}{\partial\vpb} - 4 V\biggr) 
\end{equation}
is a new function depending on the specific form of the potential. Notice 
that $g(V)$ vanishes for the class of solutions described in Section 2.
Taking  the difference of Eq. (\ref{yyI}) and (\ref{ijI}),
the resulting equation is, in some way, similar to Eq. (\ref{lambdapp}):
\begin{eqnarray}
&& (\xi''  - \psi'') - \nabla^2( \xi - \psi) + 
\psi'( \vpb' + 4 {\cal F} - 3 {\cal H}) - \xi' ( \vpb' + 3 {\cal H} - 2 {\cal F}) = 
\nonumber\\
&&-({\cal F} - {\cal H})[ E''' - (\vpb' + {\cal H}) E'' - (\vpb'' + {\cal H}') E' - 
\nabla^2 E'].
\label{xipsi}
\end{eqnarray}
Hence, combining  Eqs. (\ref{lambdapp}) and (\ref{xipsi}) 
to get rid of the terms containing the derivatives of $E$,
the following simple relation appears 
\begin{equation}
\nabla^2 \xi- \xi'' + (\vpb' + {\cal H}) \xi' +\frac{4}{3} \biggl( \frac{{\cal F} - {\cal H}}{\vpb' + {\cal H}}\biggr) g(V)\phi= \frac{2}{3} \biggl( \frac{\vpb' + 3 {\cal H} 
-2 {\cal F}}{\vpb' + {\cal H}}\biggr) [ \nabla^2 \lambda - \lambda'' + (\vpb' + {\cal H}) \lambda'].
\label{xilambda}
\end{equation}
This equation can be used either to eliminate 
the Laplacian of $\xi$ in favour of a Laplacian 
of $\lambda$ or vice-versa.

To get a quasi-decoupled equation for $\lambda$,
the first step is to take the conformal time derivative of Eq. (\ref{lambdaprime}).
Terms like $\nabla^2 E''$ will naturally appear and their 
presence can be eliminated by means of Eq. (\ref{int2}) or (\ref{Eprime}).
As a result of this manipulation terms proportional to $\nabla^2 \xi$ 
and to $\nabla^2 E'$ will arise. They can both be eliminated 
by the use of Eqs. (\ref{xilambda}) and (\ref{lambdaprime}).
The final result for the evolution equation of $\lambda$ reads then
\begin{eqnarray}
&& \lambda'' - \biggl[ 1 + \frac{3}{2} a^2 \frac{\partial V}{\partial \vpb} 
\frac{1}{(\vpb' + 3 {\cal H} + {\cal F})^2}\biggr] \nabla^2 - 
\lambda' \biggl[ ( \vpb' + {\cal H} ) + \frac{6 f(V) ( 3 {\cal H}^2 + {\cal F}^2 
{\cal H} \vpb')}{(\vpb' + {\cal H}) ( \vpb' + 3 {\cal H} + {\cal F})^2}\biggr]
\nonumber\\
&&=\frac{( {\cal F} - {\cal H})( \vpb' + 3 {\cal H} 
- 2 {\cal F})}{( \vpb' + 3 {\cal H}+ {\cal F})^2} g(V) \phi
+ \frac{3}{2} \frac{{\cal F} - {\cal H}}{(( \vpb' + 3 {\cal H} + {\cal F})^2} a^2 \frac{\partial V}{\partial \vpb} \xi'
\nonumber\\
&&+ 3 \frac{ ({\cal F} - {\cal H})( 3 {\cal H}^2 + {\cal F}^2 + {\cal H} \vpb')}{(\vpb' 
+ {\cal H})( \vpb' + 3 {\cal H} + {\cal F})^2} f(V) \xi.
\label{lamppfin}
\end{eqnarray}
As previously remarked, in the limit $\eta\to -\infty$ the potential 
terms are negligible with respect to the other terms of comparable
dimensions such as $ {\vpb'}^2$, ${\cal H}^2$, ${\cal F}^2$ and 
their mutual combinations. Consequently, the evolution equation 
for $\lambda$ becomes, in this limit
\begin{equation}
\lambda'' + 2 \frac{z_{\rm s}'}{z_{\rm s}} \lambda' - \nabla^2 \lambda =0,
\label{NM1}
\end{equation}
where 
\begin{equation}
z_{\rm s} = - 2 \sqrt{\frac{2}{3}}\frac{e^{-\vpb/2}}{\sqrt{a} } \frac{ \vpb' + 3 {\cal H} + {\cal F}}{\vpb' + {\cal H}}.
\end{equation}
Notice, as anticipated, that if we ought to use Eq.  (\ref{lamppfin}) 
for explicit numerical integration over the whole 
range of $\eta$, poles arise for $(\vpb' + {\cal H})$. This 
confirms that second-order equations such as (\ref{NM1}) 
are mandatory for the determination of the 
initial conditions but not appropriate for the 
description of the evolution of the fluctuations.

The same type of algebra leading to Eq. (\ref{NM1}) 
can be applied to find the second (asymptotic) normal 
mode of the system.  Indeed, the variable
\begin{equation}
\theta = \biggl( \frac{\vpb' + {\cal F}}{\vpb' + 3 {\cal H} + {\cal F}}\biggr) \xi - 
\biggl( \frac{ \vpb' + 3 {\cal H} - 2 {\cal F}}{\vpb' + 3 {\cal H} + {\cal F}}\biggr)
\psi,
\label{theta}
\end{equation}
also obeys the following equation
\begin{eqnarray}
 && \theta'' - (\vpb' + {\cal H}) \theta' - \nabla^2 \theta =
 \nonumber\\
 && \frac{ 2 V a^2}{(\vpb' + {\cal H}) ( \vpb' + 3 {\cal H} + {\cal F})} [ 2 ( {\cal F} 
 - {\cal H}) \lambda' - ( 2 {\cal H} + {\cal F}) \theta']
 \nonumber\\
 &&- \frac{ V a^2 ( 2 {\cal H} + {\cal F})}{(\vpb' + 3 {\cal H} + {\cal F}) ( \vpb' + 
 {\cal H})^2}[ {\vpb'}^2  - {\cal H}^2 - 2 V a^2] \theta,
 \end{eqnarray}
 whose specific form becomes, in the limit $\eta \to -\infty$, 
 \begin{equation}
 \theta'' + 2 \frac{z_{\rm s}'}{z_{\rm s}} \theta' - \nabla^2 \theta =0.
 \end{equation}
 Hence the canonical normal modes are 
 \begin{equation}
 v = z_{\rm s} \lambda,\,\,\,\,\,\,\,\,\,\,\,\,w = z_{\rm s} \theta
 \end{equation}
 and their Lagrangian density can be written as 
 \begin{equation}
 {\cal L}^{(2)} = \frac{1}{2} \biggl[ {v'}^2 + \frac{z_{\rm s}''}{z_{\rm s}} v^2 - 
 (\partial_{i} v)^2 + {w'}^2 + \frac{z_{\rm s}''}{z_{\rm s}}  w^2 - (\partial_{i} w)^2
 \biggr].
 \label{lagrangian}
 \end{equation}
 It is relevant to stress that Eq. (\ref{lagrangian}) holds, strictly 
 speaking, only in the limit of vanishing potential (i.e. $\eta \to -\infty$)
 and agrees with the Lagrangian for the normal modes 
 of a dimensionally reduced geometry discussed in \cite{normal}.
As a side remark, we ought to conclude the present section 
by noticing that, using Eq. (\ref{lamppfin}) into Eq. (\ref{xilambda}), the  equation 
\begin{eqnarray}
&& \xi'' = \biggl\{ ( \vpb' + {\cal H}) + 2 Va^2  \frac{(\vpb' + 3 {\cal H} - 2 {\cal F})
{\cal F}}{(\vpb' + 3 {\cal H} + {\cal F}) [ (\vpb' + 3 {\cal H})^2 + 2 {\cal F}^2]}
\biggr\}\xi'
\nonumber\\
&&+ 4V a^2 \frac{(\vpb' + 3 {\cal H})( \vpb' + 3 {\cal H} - 2{\cal F})}{
(\vpb' + 3 {\cal H} + {\cal F})[ ( \vpb' + 3 {\cal H})^2 + 2 {\cal F}^2]} 
\nabla^2 \lambda
\nonumber\\
&& +  6 V a^2 \frac{({\cal F} - {\cal H})( \vpb' + 3 {\cal H} - 2 {\cal F}) ( 3 {\cal H}^2 + {\cal F}^2 + {\cal H} \vpb')} {(\vpb' + 3 {\cal H} + {\cal F})^2[ ( \vpb' + 3 {\cal H})^2 + 2 {\cal F}^2]} \xi - \nabla^2 \xi
\nonumber\\
&& + 4 V a^2 \frac{(\vpb' + 3 {\cal H} - 2 {\cal F}) ( 3 {\cal H}^2 + {\cal F}^2 + 
{\cal H}\vpb'}{(\vpb' + 3 {\cal H} + {\cal F})^2 [( \vpb' + 3 {\cal H})^2 + 2 {\cal F}^2]} \nabla^2 E'
\label{xipp}
\end{eqnarray}
can be obtained. This is the equation already anticipated in Section 4 (see 
Eq. (\ref{cosmic3})).
\renewcommand{\theequation}{6.\arabic{equation}}
\section{Numerical analysis}
\setcounter{equation}{0}

To solve numerically the evolution it is essential  to have a set 
of evolution equations whose time-dependent coefficients are 
all non-singular. The relevant  set of equations
follows from Eqs. (\ref{cosmic1}), (\ref{cosmic2}) and (\ref{cosmic3}) recalling the remark related to Eq. (\ref{zeroes}).

To solve this system numerically, quantum mechanical initial
conditions have to be imposed for $t\to -\infty$. According 
to the results of the previous section, the Fourier 
modes of  $\theta$ and 
$\lambda$ will be normalized as 
\begin{eqnarray}
&& \theta_{k} = \frac{1}{\sqrt{ 2 \omega} ( \sqrt{a} z_{\rm s})} e^{- i 
\int \omega(t') dt'},
\nonumber\\
&&\lambda_{k} = \frac{1}{\sqrt{ 2 \omega} ( \sqrt{a} z_{\rm s})} e^{- i 
\int \omega(t') dt'}. 
\label{norm1}
\end{eqnarray}
From Eq. (\ref{norm1}), the asymptotic expressions of $\dot{\theta}_{k}$ 
and of $\dot{\lambda}_{k}$ can be deduced by taking the derivative 
of the two expressions with respect to $t$.
Having determined the values of $\theta_{k}$, $\lambda_{k}$ and of their derivatives it is possible to set the initial conditions for the other 
variables. The function $\lambda_{k}$ directly appears in the system.
Recalling the explicit expressions of $\lambda_{k}$ and $\theta_{k}$
in terms of $\xi_{k}$ and $\psi_{k}$, i.e.  Eqs. (\ref{lambda}) and (\ref{theta}),
the initial conditions for $\xi_{k}$ and $\dot{{\cal P}}_{k}$ can be easily obtained:
\begin{eqnarray}
&& \xi_{k} =  \frac{2}{3}\biggl[ 
\biggl(\frac{\dot{\vpb} + 3 H + F}{\dot{\vpb} + H}\biggr) \theta_{k} + \biggl( \frac{\dot{\vpb} + 3 H - 2 F}{\dot{\vpb} + H}\biggr)\lambda_{k}\biggr] ,
\nonumber\\
&& {\cal P}_{k} = \frac{2}{3}\biggl[ 
\biggl(\frac{\dot{\vpb} + 3 H + F}{\dot{\vpb} + H}\biggr) \dot{\theta}_{k} + \biggl( \frac{\dot{\vpb} + 3 H - 2 F}{\dot{\vpb} + H}\biggr)\dot{\lambda}_{k}\biggr] 
 \nonumber\\
&& +\frac{2}{3} \frac{V[ 2 (F - H) \lambda_{k} - ( 2 H + F) \theta_{k}}{(\dot{\vpb} +H)^2}.
 \end{eqnarray}
The last quantity to be dtermined is the initial condition for ${\cal E}_{k}$ 
whose expression, in terms of the initial conditions of the other 
variables, is:
\begin{eqnarray}
&&{\cal E}_{k} = \frac{ (\dot{\vpb} + 3 H - 2 F) }{2 \omega^2 (\dot{\vpb} + H)} 
{\cal P}_{k} - \frac{( F - H) ( 3 H^2 + F^2 - \dot{\vpb}^2)}{
\omega^2(\dot{\vpb} + H)^2} \xi_{k}
\nonumber\\
&& - \frac{(\dot{\vpb}  + 3 H)^2 + 2 F^2}{\omega^2  (\dot{\vpb} + H)^2} \dot{\lambda}_{k} + \frac{2}{(\dot{\vpb} + H)} \lambda_{k}.
\end{eqnarray}

The initial time of integration, $t_{\rm i}$, is fixed in such a way that the given 
physical frequency is larger than the Hubble rate, i.e. the 
corresponding wavelength is smaller than the Hubble radius.
In the case of the class of solutions introduced in Eqs. (\ref{as1})--(\ref{phs1})
this requirement implies, for a given Fourier mode $k$,  
$t_{\rm i} < - (2 \kappa)^{-2/3} \,\,t_{0}$ where $\kappa= k t_{0}$. Thus,
as $\kappa$ becomes smaller, the initial integration time becomes 
larger in absolute value.

\begin{figure}
\centerline{\epsfxsize = 9cm  \epsffile{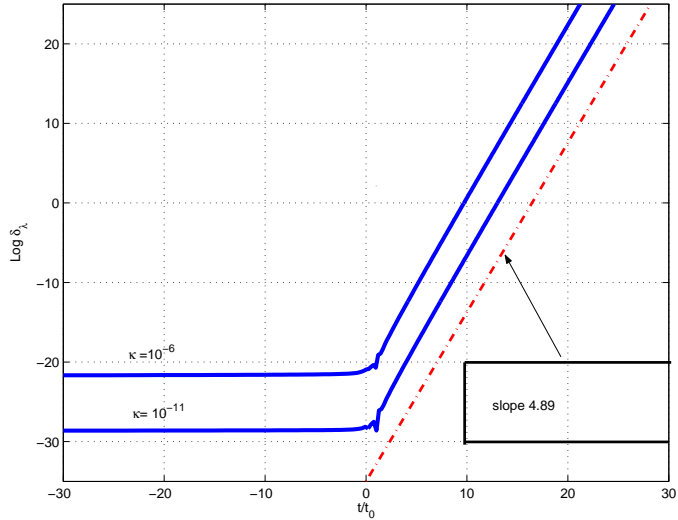}}
\vskip 3mm
\caption[a]{The evolution of the fluctuations is illustrated in the case of a 
five-dimensional background belonging to the class (\ref{as1})--(\ref{phs1}).
We took $\varphi_{0}\sim -40$. As previously 
introduced $\kappa = k\, t_{0}$. }
\label{FIG1} 
\end{figure}

Consider now the five-dimensional background geometry obtained 
from Eqs. (\ref{as1})--(\ref{phs1}) by setting $ d= 3$ and $n=1$.
In this case Eqs. (\ref{cosmic1}), (\ref{cosmic2}) and (\ref{cosmic3}) 
can be integrated by setting quantum mechanical initial 
conditions as discussed in the previous paragraphs of the present 
section. The result of the numerical integration is illustrated in 
Fig. \ref{FIG1} where the logarithm (in ten basis) of 
spectrum of $\lambda$, i.e. 
\begin{equation}
\delta_{\lambda} = k^{3/2} |\lambda_{k}|^2,
\label{spectrum}
\end{equation}
is reported. In analogy with Eq. (\ref{spectrum}) we will also 
define, in similar terms, the spectrum of $\xi$, i.e. $\delta_{\xi}$ 
and so on. 

It is clear from Fig. \ref{FIG1} as well as from 
the equivalent plots obtained for $\delta_{\xi}$ and $\delta_{{\cal E}}$
that, in spite of the initial conditions, for $t>0$ the following 
approximate relation holds 
\begin{equation}
\delta_{\lambda}(t) \sim \delta_{\xi}(t)\sim \delta_{{\cal E}}(t) \sim A(k) e^{ \alpha (t/t_{0})}
\label{RES1}
\end{equation}
where $\alpha$ is a constant and it is 
numerically determined to be $4.9$; the symbol of similarity in Eq. (\ref{RES1}) means that the  relation 
holds up to constant factors of order $1$.  This occurrence signals that there is 
a kind of late-time attractor in Eqs. (\ref{cosmic1}), (\ref{cosmic2}) and (\ref{cosmic3}).
 According to Eqs. (\ref{RES1}), the plots 
for the other power spectra will be quantitatively and qualitatively similar to the one of $\lambda$. The sharp (exponential) amplification of the fluctuations, is rather dangerous for the validity of the perturbative expansion, as 
it can be argued from Fig. \ref{FIG1}.
\begin{figure}
\centerline{\epsfxsize = 9cm  \epsffile{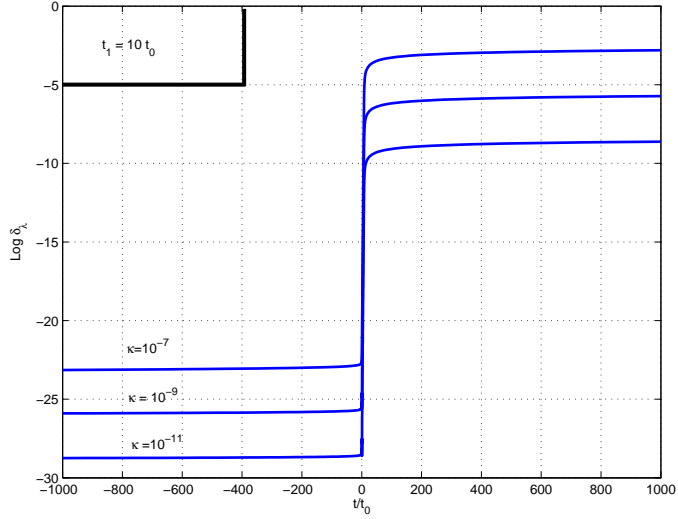}}
\vskip 3mm
\caption[a]{The evolution of curvature fluctuations is illustrated in the case 
where $F\to 0$ for $t>t_{1}$ with $t_{1}= 10$.}
\label{FIG2} 
\end{figure}

To gain analytical understanding of the behaviour
reported in Fig. \ref{FIG1} let us write the system provided by 
Eqs. (\ref{cosmic1}), (\ref{cosmic2}) and (\ref{cosmic3}) 
in the limit of large (positive) times, i.e. $t \gg t_{0}$. Denoting 
with $\tau= t/t_{0}$ and with $\kappa = k t_{0}$ the system 
becomes:
\begin{eqnarray}
&& \dot{\lambda}_{k} \simeq \biggl( \frac{2}{3 \tau^2} - \frac{1}{2}\biggr) {\cal P}_{k}
+ \biggl( \frac{4}{3 \tau^3} - \frac{26}{9 \tau^{5}}\biggr) \xi_{k} + \biggl(\frac{17}{18}\frac{\kappa^2}{\tau^2}- \frac{2}{3} \kappa^2\biggr) \lambda_{k} + \biggl( \frac{35}{72} \frac{\kappa^2}{\tau^3} - 
\frac{\kappa^2}{6 \tau} \biggr) {\cal E}_{k},
\nonumber\\
&& \dot{{\cal E}}_{k} \simeq \biggl(\frac{2}{3} \kappa^2 - \frac{17}{18} \frac{\kappa^2}{\tau^2} \biggr) {\cal E}_{k} + \biggl( 2 \tau + \frac{1}{3 \tau}
\biggr){\cal P}_{k} + \biggl( \frac{3}{2} - \frac{10}{3 \tau^2}\biggr) \xi_{k} + 
\biggl( \frac{8}{3} \kappa^2 \tau + \frac{2 \kappa^2}{9 \tau} -1 \biggr) 
{\cal E}_{k},
\nonumber\\
&& \dot{\xi}_{k} \simeq {\cal P}_{k},
\nonumber\\
&& \dot{{\cal P}}_{k} \simeq \biggl( \frac{3}{\tau} - \frac{10}{3 \tau^3}\biggr) {\cal P}_{k}  +\biggl( 8 \kappa^2 \tau + \frac{22 \kappa^2}{3 \tau} \biggr) \lambda_{k}
+ \biggl( 24 + \frac{\kappa^2}{2 \tau}\biggr) \xi_{k} + \biggl( 8 \kappa^2 - 
\frac{ 2 \kappa^2}{ 3 \tau^2} \biggr) {\cal E}_{k}.
\label{CE}
\end{eqnarray}
In Eq. (\ref{CE}) we kept all the leading terms in the limit $\tau\gg 1$ 
as well as the first subleading correction. The system (\ref{CE}) 
can be solved in the limit $\kappa \to 0$, i.e. when all the 
modes have wavelengths larger than the Hubble radius at $t_{0}$.
In this limit the system becomes
\begin{equation}
\ddot{\xi}_{k} \simeq 24\,\, \xi_{k}, \,\,\,\,\,\,\,\,\,{\cal P}_{k} \simeq \dot{\xi}_{k},
\,\,\,\,\,\,\,\,\,\dot{{\cal E}}_{k} \simeq \frac{3}{2} \,\,\xi_{k},\,\,\,\,\,\,\,\,\,\,
\dot{\lambda}_{k} \simeq -\frac{1}{2}\, {\cal P}_{k}.
\label{CE2}
\end{equation}
The system (\ref{CE2}) can be easily 
solved and the result is that an exponential mode
\begin{equation}
\xi_{k}(t) \simeq \xi_{0}(k)\,\, e^{2\sqrt{6} \tau}, \,\,\,\,\,\,\,\,\, 
\lambda_{k}(t) \simeq - \sqrt{6}\xi_{0}(k)\,\, e^{ 2 \sqrt{6} \tau},
\,\,\,\,\,\,\,\,\, {\cal E}_{k}(t) \simeq \frac{\sqrt{3}}{4 \sqrt{2}} \,\,\xi_{0}(k)\,\,
e^{2 \sqrt{6} \tau},
\label{CE3}
\end{equation}
may arise.
Equation (\ref{CE3}) is the rationale for the result of Eq. 
(\ref{RES1}) and, indeed, $ 2\sqrt{6} \simeq 4.8989$. 
In Fig. \ref{FIG1},
the dot-dashed line is nothing but the slope 
of the logarithm (in ten basis) of the function $e^{2 \sqrt{6} (t/t_{0})}$. 
This example shows that the fluctuations of the internal
dimensions provide an effective non-adiabatic pressure 
variation whose r\^ole is to provide a source term for 
curvature perturbations. As it is evident from this 
example, the continued contraction of the internal
scale factor is not sufficient for a perturbative 
description of the fluctuations. In other words, it is 
the fact that $F$ does not vanish for $t> t_{0}$ 
that triggers the exponential growth of the 
curvature fluctuations.  

To stress this point, let us now consider 
the case when the solution is exactly the one previously 
studied but only for $t < t_{1}$. For $t> t_{1}$ the 
Universe is described by a five-dimensional geometry 
where, however, $b\sim constant$ and $F$ is driven dynamically 
to zero. Under this assumption let us now solve the evolution 
equations of the fluctuations, i.e. Eqs. (\ref{cosmic1}), (\ref{cosmic2}) 
and (\ref{cosmic3}). Of course, as discussed above in the present and in the 
previous section, quantum mechanical normalization will be enforced 
for $ t\to - \infty$.  
In Fig. \ref{FIG2} the result of such an exercise is reported in the case 
$t_1=10 \,\,t_{0}$. It is clear that the dangerous exponential amplification 
is absent and that the treatment of the inhomogeneities 
is perturbative while the relevant wavelengths are larger 
than the Hubble radius.
\begin{figure}
\centerline{\epsfxsize = 9cm  \epsffile{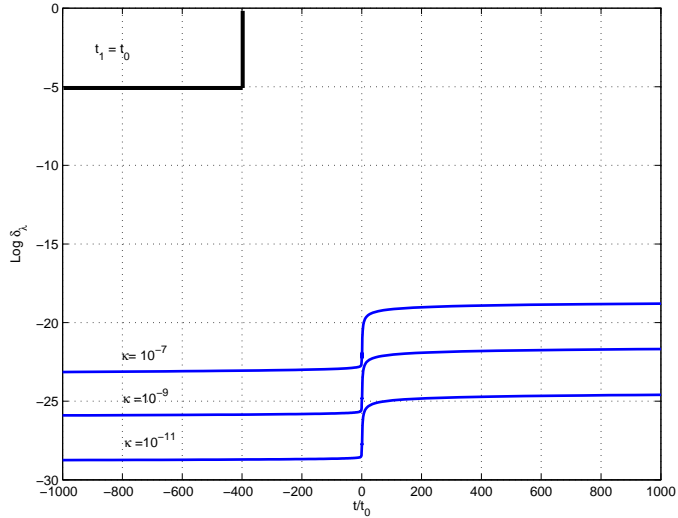}}
\vskip 3mm
\caption[a]{The same analysis discussed in Fig. \ref{FIG2}
is here repeated in the case $t_{1} = t_{0}$.}
\label{FIG3} 
\end{figure}
If $b(t)$ freezes at an earlier epoch, the amplitude 
of the curvature fluctuations will be smaller. This 
observation is corroborated by Fig. \ref{FIG3} 
where the same quantity illustrated in Fig. \ref{FIG2}
is computed in the case $t_{1} \simeq t_{0}$. 
Since $t_{1}$ is smaller, $F\to 0$ earlier than in the 
case of Fig. \ref{FIG2}. This implies that the (short) phase 
of exponential amplification present 
in the example of Fig. \ref{FIG2} is now absent.

\renewcommand{\theequation}{7.\arabic{equation}}
\section{Concluding remarks}
\setcounter{equation}{0}

In the present paper the possible occurrence of dimensional 
decoupling has been scrutinized in light of the possible 
space-time fluctuations of the internal dimensions. 
Dimensional decoupling arises in various classes 
of higher-dimensional (anisotropic) manifolds 
characterized by the expansion of the observable dimensions 
and by the contraction of the internal dimensions. In the 
present paper, dimensional decoupling has been 
described by means of a class of solutions of the low-energy 
string effective action. The virtues of this description are that 
the background solutions are always regular. Therefore, a
reasonably accurate description of the evolution of the inhomogeneities 
can be developed. 

Various results have been obtained:
\begin{itemize}
\item{} the full system of the equations for the perturbations has been derived
 in the string frame description;
\item{} the evolution equations of the zero-modes 
have been diagonalized in terms of a set of variables
that are non-singular throughout the evolution 
of the background geometry;
\item{} quasi-normal modes of the system have found;
\item{} a numerical analysis of the appropriate (first-order)
evolution equations has been presented.
\end{itemize}

The obtained results show
that the contraction of the internal dimensions 
is not a sufficient requirement for the perturbative 
description of the inhomogeneities. If the scale factor of the internal dimensions is not fixed to a constant (asymptotic) value, then 
the evolution of the internal fluctuations triggers  the presence 
of a source term for the evolution of curvature fluctuations. This 
term may be interpreted as the analog of a non-adiabatic
pressure-density fluctuations whose r\^ole, in four space-time 
dimensions, is known to imply a growth of curvature fluctuations. 
A possible problem arises, in this context, if the contraction 
of the internal manifold lasts for a long time. In this 
case the numerical results show that the evolution equations of the 
fluctuations may develop some kind of dynamical 
attractor where the curvature fluctuations are 
exponentially amplified in spite of the initial conditions. 
The numerical analysis can be corroborated, in the specific 
case under consideration, by a precise analytical understanding.
This potential problem of dimensional decoupling 
can be avoided provided the internal scale 
factor sets, sufficiently early, to a constant value. In this 
case the evolution of the fluctuations may still be perturbative 
during the whole evolution.

The results reported in the present paper suggest that realistic models of multidimensional
pre-big bang dynamics can be constructed. However, 
unlike the attitude taken so far,
care must be taken in dealing with the fluctuations 
of the internal dimensions that can induce dangerous
late-time attractors in the evolution equations 
of the fluctuations. 
\newpage 

\begin{appendix}
\renewcommand{\theequation}{A.\arabic{equation}}
\setcounter{equation}{0}
\section{Technical details}
In the case of the geometry given in Eq. (\ref{geometry}), assuming that 
both the external and internal manifolds are spatially flat, the Christoffel 
connections can be written as 
\begin{eqnarray}
&&\Gamma_{00}^{0} = {\cal H}, ~~~~\Gamma_{i j}^{0} = {\cal H} \delta_{ij},~~~~~
\Gamma_{0j}^{i} = {\cal H} \delta_{i}^{j},
\nonumber\\
&& \Gamma_{a b}^{0} = \frac{b^2}{a^2}{\cal F} \delta_{a b},~~~~
\Gamma_{a 0}^{b} = {\cal F} \delta_{a}^{b}, 
\end{eqnarray}
while the components of the Ricci tensor are:
\begin{eqnarray}
&& R_{0}^{0} = - \frac{1}{a^2} [ d {\cal H}' + n{\cal F}' + n{\cal F}^2 - n {\cal H} {\cal F}],
\nonumber\\
&& R_{i}^{j} = -\frac{1}{a^2} [ {\cal H}' + (d-1) {\cal H}^2 + n {\cal H} {\cal F}] 
\delta_{i}^{j},
\nonumber\\
&& R_{a}^{b} = - \frac{1}{a^2} [ {\cal F}' + (d-1) {\cal H}{\cal F} + n{\cal F}^2] 
\delta_{a}^{b}.
\label{RICCI}
\end{eqnarray}
From Eq. (\ref{RICCI}), the  Ricci scalar is then 
\begin{equation}
R= - \frac{1}{a^2}[ 2 d {\cal H}' + 2 n {\cal F}' + n(n + 1) {\cal F}^2 
+ d (d -1) {\cal H}^2 + 2 n (d -1) {\cal H}{\cal F}].
\end{equation}
In the uniform dilaton gauge (see  Eq. (\ref{gaugecond})) the 
non-vanishing components of the perturbed metric (\ref{param}) become, 
\begin{eqnarray}
&& \delta G_{00} = 2 a^2 \phi,
\nonumber\\
&& \delta G_{0y} = - a b C,
\nonumber\\
&&\delta G_{i j} = 2 a^2 \psi \delta_{i j} - 2 a^2 \partial_{i} \partial_{j} E,
\nonumber\\
&& \delta G_{yy} = 2 b^2 \xi
\end{eqnarray}
and 
\begin{eqnarray}
&& \delta G^{00} =  -\frac{2\phi}{a^2},
\nonumber\\
&& \delta G^{0y} = - \frac{C}{ab},
\nonumber\\
&&\delta G^{i j} = -\frac{2  \psi}{a^2} \delta_{i j} + 
\frac{2  \partial^{i} \partial^{j} E}{a^2},
\nonumber\\
&& \delta G^{yy} = -\frac{2 \xi}{b^2}.
\end{eqnarray}
The fluctuations of the  Christoffel connections become then
\begin{eqnarray}
&& \delta \Gamma_{0 0}^{0} = \phi',
\nonumber\\
&& \delta \Gamma^{i}_{00} = \partial^{i} \phi,
\nonumber\\
&& \delta \Gamma_{i j}^{0} = - [ \psi'\delta_{ij} + 2 {\cal H} ( \phi + \psi)] \delta_{i j} + 
\partial_{i} \partial_{j} [ E' + 2 {\cal H} E],
\nonumber\\
&& \delta \Gamma_{i 0}^{0} = \partial_{i} \phi,
\nonumber\\
&& \delta\Gamma_{i0}^{j} = - \psi' \delta_{i}^{j} + \partial_{i} \partial^{j} E'
\nonumber\\
&& \delta \Gamma_{yy}^{0} = - \frac{b}{a} \partial_{y} C - \frac{b^2}{a^2} [ \xi' + 2 {\cal F}(\phi + \xi)],
\nonumber\\
&& \delta \Gamma_{0 y}^{y}= - \xi',
\nonumber\\
&& 
\delta\Gamma_{00}^{y} = \frac{a^2}{b^2} \partial^{y} \phi + \frac{a}{b} [ C' + {\cal F} C],
\nonumber\\
&& \delta\Gamma_{y 0}^{0} = \partial_{y} \phi + \frac{b}{a} {\cal F} C,
\nonumber\\
&& \delta \Gamma_{yy}^{y} = - \partial_{y} \xi - \frac{b}{a} {\cal F} C,
\nonumber\\
&& \delta \Gamma_{y y}^{i} = \frac{b^2}{a^2} \partial^{i} \xi,
\nonumber\\
&& \delta\Gamma_{i y}^{y} = \partial_{i} \xi,
\nonumber\\
&& \delta\Gamma_{y i}^{0} = -\frac{b}{2 a} \partial_{i} C,
\nonumber\\
&&\delta\Gamma_{i 0}^{y} = \frac{a}{2 b} \partial_{i} C,
\nonumber\\
&& \delta\Gamma_{0y}^{i} =- \frac{b}{2 a} \partial^{i} C,
\nonumber\\
&& \delta\Gamma_{i j}^{y} = - \frac{a}{b} {\cal H} C \delta_{i j} + 
\frac{a^2}{b^2}  \partial^{y} [\psi \delta^{i j} - \partial_{i} \partial_{j} E],
\nonumber\\
&& \delta \Gamma_{i y}^{j} = - \partial_{y} \psi \delta_{i}^{j} + \partial_{y} \partial_{i}\partial^{j} E ,
\nonumber\\
&& \delta\Gamma_{i j}^{k} = 
\partial^{k} \psi \delta_{i j} - \partial_{j}\psi \delta^{k}_{i} 
-\partial_{i} \psi \delta_{j}^{k} + \partial_{i}\partial_{j}\partial^{k} E.
\end{eqnarray}
The first order fluctuations of the  Ricci tensor with mixed indices are:
\begin{eqnarray}
&&\delta R_{0}^{0} = \frac{1}{a^2} \nabla^2[ \phi - E'' - {\cal H} E' ] + 
\frac{1}{b^2} \nabla^{2}_{y} \phi + 
\nonumber\\
&&\frac{1}{a b}\partial_{y}( C' + {\cal F} C) + \frac{1}{a^2} [ 3 \psi'' + \xi'' + 
( 3 {\cal H} \psi' + ( 2 {\cal F} - {\cal H}) \xi') ]
\nonumber\\
&&+ 
\frac{1}{a^2}( 6 {\cal H}' + 2 {\cal F}' - 2 {\cal H} {\cal F} + 2 {\cal F}^2) ,
\nonumber\\
&&\delta R_{0}^{i} = - \frac{1}{a^2} \partial^{i}[ 2 \psi' + \xi' + \xi ( {\cal F} - {\cal H})
+ (2 {\cal H} + {\cal F}) \phi  + \frac{b}{2 a} C],
\nonumber\\
&& \delta R_{i}^{j} = \frac{1}{a^2} \delta_{i}^{j} \{ \psi'' + [ 2 {\cal H}' + 4 {\cal H}^2 + 
2{\cal H} {\cal F} ]\phi + {\cal H} \phi' 
\nonumber\\
&& + (5 {\cal H} + {\cal F}) \psi' + 
{\cal H}\xi' - \nabla^2 \psi - \frac{a^2}{b^2} \nabla^{2}_{y} \psi - 
{\cal H} \nabla^2 E' + \frac{a}{b} {\cal H} \partial_{y} C\}
\nonumber\\
&&+ \partial_{i}\partial^{j}[ - E'' - ( 2 {\cal H} + {\cal F}) E' + \phi - \xi - \psi + 
\frac{a^2}{b^2} \nabla^{2}_{y} E ],
\nonumber\\
&& \delta R_{y}^{y} = \frac{1}{a^2} \{ \xi'' + 2 ({\cal H} + {\cal F}) \xi' + {\cal F}\phi' 
+ 3 {\cal F} \psi' + \phi ( 2 {\cal F}^2 + 4 {\cal H} {\cal F} + 2 {\cal F}')
- \nabla^2 \xi - {\cal F} \nabla^2 E'\} 
\nonumber\\
&&+ \frac{1}{b^2} \nabla^{2}_{y}[\phi + \nabla^2 E - 3 \psi]+ 
\frac{1}{a b} \partial_{y}[ C' + (3 {\cal H} + {\cal F}) C],
\nonumber\\
&& \delta R_{0}^{y} = \frac{C}{a b} [ 3 {\cal H}' - 2 {\cal H}{\cal F}] + 
\frac{1}{2 a b} \nabla^{2} C 
\nonumber\\
&&- 
\frac{1}{b^2} \partial^{y} \{ 3 \psi' + 3 {\cal H} \phi + 3 ( {\cal H} - {\cal F}) \psi 
- \nabla^2[E' + ({\cal H} - {\cal F}) E]\},
\nonumber\\
&& \delta R_{i}^{y} =  \frac{1}{2 a b} \partial_{i}[ C' + ( {\cal H} + 2 {\cal F})C] -
\frac{1}{b^2} \partial_{i}\partial^{y}[ 2 \psi - \phi],
\end{eqnarray}
while the fluctuation of the Ricci scalar becomes:
\begin{eqnarray}
&& \delta R =  \frac{2}{a^2} \biggl\{ \nabla^2[ \phi - \xi - 2 \psi - E'' - 
( 3 {\cal H} + {\cal F}) E' + \frac{a^2}{b^2} \nabla^2_{y} E]
\nonumber\\
&&+ 3 \psi'' + \xi'' + (3 {\cal H} + {\cal F}) \phi' + 
( 9 {\cal H} + 3 {\cal F}) \psi' + 2 \xi' ( {\cal H} + {\cal F}) 
\nonumber\\
&&+
(6 {\cal H}' + 2 {\cal F}' + 4 {\cal H}{\cal F} + 6 {\cal H}^2 + 2 {\cal F}^2 ) \phi
\nonumber\\
&& + \frac{2}{b^2} \nabla^2_{y}( \phi - 3\psi) + 
\frac{2}{a b} \partial_{y}[ C' + ( 3 {\cal H} + {\cal F}) ] .
\end{eqnarray}
Defining now 
\begin{equation}
{\cal G}_{A}^{B} = \delta R_{A}^{B} - \frac{1}{2} \delta_{A}^{B} \delta R,
\end{equation}
the scalar fluctuation of the Einstein tensor leads to
the following components:
\begin{eqnarray}
&& \delta {\cal G}_{0}^{0} = \frac{1}{a^2} \nabla^2[ \xi + 2 \psi + ( 2 {\cal H} + {\cal F}) E'
- \frac{a^2}{b^2}\nabla^2_{y} E]
\nonumber\\
&& - \frac{3}{a b} {\cal H} \partial_{y} C + \frac{3}{b^2} \nabla^2_{y} \psi 
+ \frac{1}{a^2} [ - 3 ( 2 {\cal H} + {\cal F}) \psi' - 3 {\cal H} \xi' 
- 6 {\cal H} ( {\cal H} + {\cal F}) \phi],
\nonumber\\
&&\delta {\cal G}_{i}^{j} = \frac{1}{a^2} \delta_{i}^{j}\biggl\{ - 2 \psi'' - \xi'' 
-2 [ {\cal H}^2 + {\cal F}^2 + {\cal H}{\cal F} + {\cal F}' + 2 {\cal H}' ] \phi 
\nonumber\\
&& - ({\cal F} + 2 {\cal H})\phi'  - 
2 ({\cal F} + 2 {\cal H}) \psi' - ({\cal H} + 2 {\cal F}) \xi' 
\nonumber\\
&& + \nabla^2[ E'' + ( 2 {\cal H} + {\cal F}) E' + \psi + \xi - \phi - \frac{a^2}{b^2} 
\nabla^{2}_{y} E]\biggr\}
\nonumber\\
&& + \frac{\delta_{i}^{j}}{b^2} \nabla^2_{y}( 2\psi - \phi) - \frac{1}{a b}\partial_{y}[ C' 
+ (2 {\cal H} + {\cal F}) ]\delta_{i}^{j}
\nonumber\\
&&- \frac{1}{a^2} \partial_{i} \partial^{j}[E'' + ( 2 {\cal H} + {\cal F}) E' + \psi + \xi - \phi - \frac{a^2}{b^2} 
\nabla^{2}_{y} E] ,
\nonumber\\
&& \delta {\cal G}_{y}^{y} = \frac{1}{a^2} \biggl\{ -3 \psi'' - 3 {\cal H} \phi' - 9 {\cal H}\psi' - 6 ({\cal H}' + {\cal H}^2) \phi 
\nonumber\\
&& + \nabla^2[ 2 \psi - \phi + E'' + 3 {\cal H} E']\biggr\},
\label{EINT}
\end{eqnarray}
where, clearly, $\delta {\cal G}_{0}^{i} = \delta R_{0}^{i}$, 
$\delta {\cal G}_{0}^{y} = \delta R_{0}^{y}$,  $\delta {\cal G}_{i}^{y} = \delta R_{i}^{y}$.

  Finally, the general form of the perturbed equations can be usefully expressed 
in a compact way as
\begin{eqnarray}
&& a^2 \delta {\cal G}_{0}^{0} + \phi[ 2 {\cal F} {\vp}' - {{\vp}'}^2 + 6 {\cal H}{\vp}']  
\nonumber\\
&&+ 3 {\vp}' \psi' + {\vp}' \xi' - {\vp}' \nabla^2 E' + \frac{a}{b} {\vp}' \partial_{y} C =0,
\nonumber\\
&&  a^2 \delta {\cal G}_{i}^{j} +\vp' \partial_{i}\partial^{j}E' 
+ \delta_{i}^{j}\biggl\{ \phi[ 2 {\vp}'' - {{\vp}'}^2 + 2 {\cal H} {\vp}' + 
2 {\cal F} {\vp}' - \frac{a^2}{2} \frac{\partial V}{\partial \vpb}] 
\nonumber\\
&& + 2 \varphi' \psi' + 
\varphi' \xi' - \varphi' \nabla^2 E' + \varphi' \phi' + \frac{a}{b} \varphi' \partial_{y} C
\biggr\} =0,
\nonumber\\
&& a^2 \delta {\cal G}_{y}^{y} + \phi [ 2 {\vp}'' - {{\vp}'}^2 + 4 {\cal H} {\vp}' - 
\frac{a^2}{2} \frac{\partial V}{\partial \vpb} ]+ \phi' \vp' + 3 \vp' \psi' - 
{\vp}' \nabla^2 E' =0,
\nonumber\\
&&  a^2 \delta {\cal G}_{0}^{y} + \frac{a}{b} C [ - {\vp}'' + ( {\cal H} + {\cal F}) {\vp}' + 
\frac{a^2}{2} \frac{\partial V}{\partial \vpb} ] + \varphi' \frac{a^2}{b^2} \partial^{y} 
\phi=0,
\nonumber\\
&& a^2 \delta {\cal G}_{i}^{y}  - \frac{a}{2 b} {\vp}' \partial_{i} C =0,
\nonumber\\
&& a^2 \delta {\cal G}_{0}^{i} + \vp' \partial^{i} \phi  =0.
\label{EOM}
\end{eqnarray} 
Equations (\ref{EINT}) together with Eqs. (\ref{EOM}) 
can be used to derive the evolution equations 
of the fluctuations discussed in the bulk of the present paper.
\end{appendix}

\end{document}